\documentclass[]{spie}  
\usepackage[]{graphicx}
\usepackage{amsmath,amssymb}

\newcommand{\eq}{\begin{equation}}
\newcommand{\en}{\end{equation}}
\newcommand{\eqa}{\begin{eqnarray}}
\newcommand{\ena}{\end{eqnarray}}

\def\H{{\cal H}}
\def\ra{\rangle}

\newcommand{\ket}[1]{|#1\rangle}

\newsavebox{\fmbox}
\newenvironment{fmpage}[1]
     {\medskip\begin{lrbox}{\fmbox}\begin{minipage}{#1}}
     {\end{minipage}\end{lrbox}\fbox{\usebox{\fmbox}}\medskip}

\title{Post-Quantum Key Exchange Protocols}

\author{Xiangdong Li\supit{a}, Lin Leung\supit{b}, Andis Chi-Tung
Kwan\supit{c}, Xiaowen Zhang\supit{c}, Dammika Kahanda\supit{c},
Michael Anshel\supit{d} \footnote{\,\,\,\,Students from
\emph{Quantum Computing} and \emph{Quantum Cryptography} classes at
the CUNY Graduate Center made contributions to this study.}
\skiplinehalf \supit{a} CST, NYC College of Technology, CUNY, 300
Jay Street, Brooklyn, NY 11201 \skiplinehalf \supit{b} Computer
Information Science, BMCC, CUNY, 199 Chambers Street, NY, NY 10007
\skiplinehalf \supit{c} Computer Science, Graduate Center, CUNY, 365
$5^{\mbox{th}}$ Ave, NY, NY 10016 \skiplinehalf
\supit{d} Computer Science, CCNY, CUNY, $138^{\mbox{th}}$ Street, NY, NY 10031 \\
}

\authorinfo{Further information please contact:\\XiangDong Li:
E-mail:xli@citytech.cuny.edu \\ Michael Anshel:
E-mail:mikeAt1140@aol.com}

\begin{document}
  \maketitle

\setlength{\baselineskip}{10pt} {
\begin{abstract}
If an eavesdropper Eve is equipped with quantum computers, she can
easily break the public key exchange protocols used today. In this
paper we will discuss the post-quantum Diffie-Hellman key exchange
and private key exchange protocols.
\end{abstract}



\section{Why Post-Quantum Key Exchange?}
\label{sect:intro}  

Diffie and Hellman proposed the first public-key agreement for key
exchange in 1976. This protocol relies on the difficulty of
computing discrete logarithms in a finite field. The most popular
public key algorithm for encryption and digital signature is RSA.
The security of RSA is based on the intractability of the integer
factorization problem. There are a few other cryptographic schemes
that are used in practice, for example, the Digital Signature
Algorithm (DSA) and the Elliptic Curve Digital Signature Algorithm
(ECDSA). The security of these schemes is based on the discrete
logarithm problem in the multiplicative group of a prime field or in
the group of points of an elliptic curve over a finite field.

But in 1994 Shor \cite{Shor97} showed that quantum computers can
break all digital signatures that are used today. In 2001 Chuang et
al \cite{Vandersypen01} implemented Shor's algorithm on a $7-qubit$
quantum computer. When quantum computers reach approximately $30$ to
$40$ $q-bits$ they will start to have the speed (parallelism) needed
to attack the methods society uses to protect data and processes,
including encryption, digital signatures, random number generators,
key transmission, and other security algorithms.

We cannot predict exactly when this will happen because each advance
in the number of $q-bits$ has had radically different hardware
architecture. We believe quantum computers will surpass the speed of
``Moore's Law" computers in the next $15$ years, break encryption in
$25$ years, and break the responding enhanced encryption (with much
longer key lengths) in $30$ to $50$ years.

Most planners don't look $20$ years into the future, and propose to
defend against quantum computer attacks by lengthening the keys.
However, we can also defend against quantum computer attacks by
researching a way which is somewhat or wholly immune to quantum
computer attacks. Many quantum public key exchange protocols have
been studied, for example BB84 and B92\cite{B92}\,. We will look at
two schemes that achieve key agreement protocol.

The heart of our key exchange protocol is to use a public satellite
-- continually broadcasting random bits at a rate so high that no
one could store more than a small fraction of them. Parties that
want to communicate in privacy share a relatively short key that
they both use to select a sequence of random bits from the public
broadcast; the selected bits serve as an encryption key for their
messages. An eavesdropper cannot decrypt an intercepted message
without a record of the random broadcasts, and cannot keep such a
record because it would be too voluminous. How much randomness would
the satellite have to broadcast? Rabin and Ding \cite{DingR02}
mention a rate of $50$ gigabits per second, which would fill up some
$800,000$ CD-ROMs per day.

The general framework is shown in Figure 1:
\begin{center}
\begin{fmpage}{15cm}
\textbf{General Key Agreement Framework}
\begin{enumerate}
  \item Random source: a satellite sends random bit signals.
  \item The two communicating parties Alice and Bob get these signals
  \item They need to know when they should count the bits as the key.
  \item Two ways: Teleportation or Quantum clock synchronization.
  \item They agree to flip one bit or more.
\end{enumerate}
\end{fmpage}
\end{center}

   \begin{figure}
   \begin{center}
   \begin{tabular}{c}
   \includegraphics [height=12cm, width=15cm] {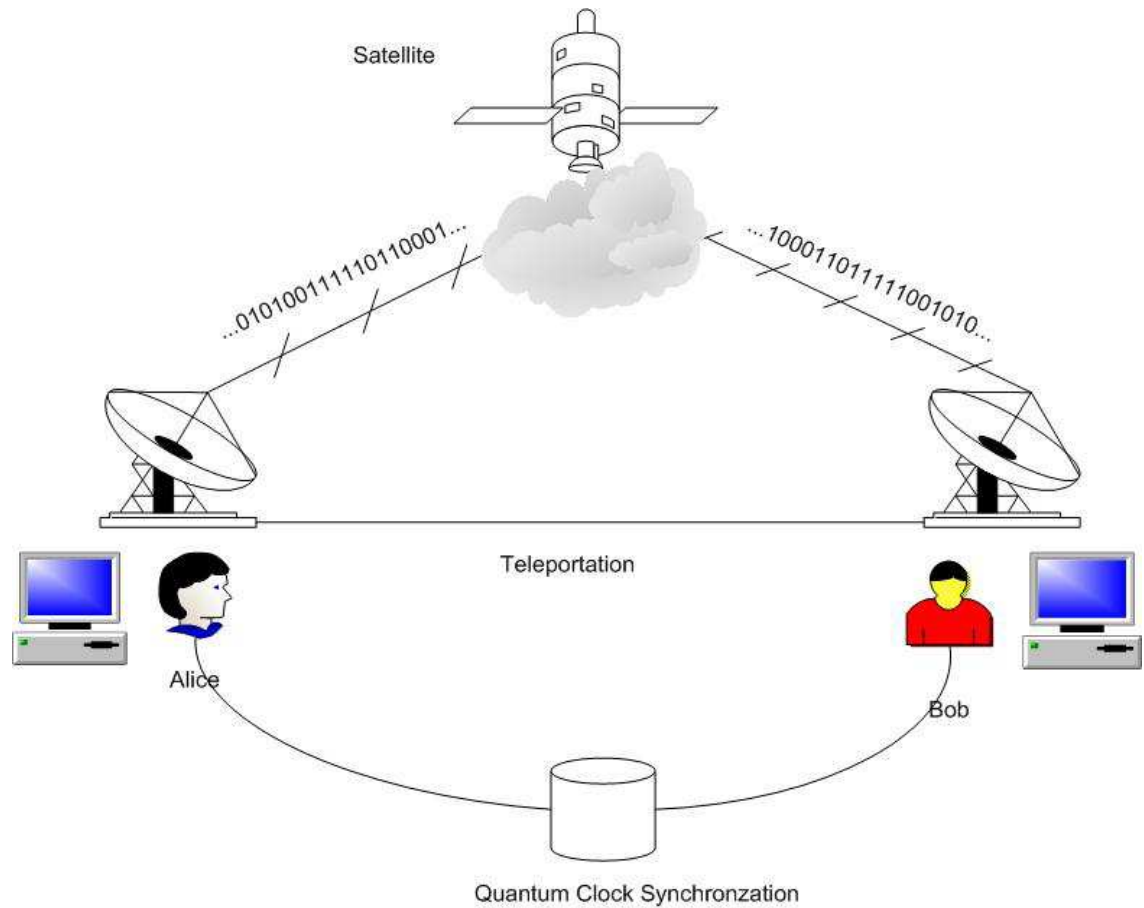}
   \end{tabular}
   \end{center}
   \caption[example]
   { \label{fig:example}
General framework of the post-quantum key exchange scheme.}
   \end{figure}

A geostationary satellite can be used as a data source generating a
random bit stream. Two communicating parties, Alice and Bob with
dish antennas, are able to receive the bit signal from the
satellite. When they want to encrypt the message, they catch the
random bits of the signal as a key. They make a public agreement on
the key size, for example, $1024$ bits.  The key is never stored in
the computer's memory, so they essentially vanish even as the
message is being encrypted and decrypted.

In order for both Alice and Bob to count the same bits as the key
from the satellite signals, three problems should be solved:

\begin{enumerate}
  \item Due to the different distances between the satellite to Alice and to
Bob, they will not count the same bits.
  \item Alice and Bob should know the starting times that they can count the same
number of bits as a key.
  \item Alice and Bob should determine the time difference between their
spatially separated clocks. For example, the determination of the
difference should be better than $100$ $ns$.
\end{enumerate}

The first problem is easily solved by using Global Position Systems
(\textbf{GPS}) to determine their positions and calculate the time
delay due to the different distance from the satellite to the
receivers. We propose to use the technology of quantum teleportation
and quantum clock synchronization to solve the latter two problems.

The organization of the paper is as follows: In Section 2 and 3, we
describe post-quantum Diffie-Hellman key, private key exchange and
quantum random walk protocols. A conclusion is given in Section 4.
We provide the fundamentals of random source, random number
generator, quantum teleportation, quantum clock synchronization, and
quantum random walk in the Appendix.

\section{Post-Quantum Key Exchange}

\subsection{Diffie-Hellman Key Exchange}
With a symmetric cryptosystem, it is necessary to transfer a secret
key to both communicating parties before secure communication can
begin. Diffie-Hellman key exchange protocol allows two parties that
have no prior knowledge of each other, to jointly establish a shared
secret key over an insecure communication channel. The first
practical scheme, Diffie-Hellman Discrete Log (see \ref{sect:dlp}
for classes of candidate) key exchange protocol, begins with two
users Alice and Bob who want to exchange two secret integers $a$ and
$b$. They agree on two public parameters, large prime $p$ and base
$g$. The protocol is specified as follows:
\begin{center}
\begin{fmpage}{15cm}
\textbf{Diffie-Hellman Key Exchange Protocols} Public announcement:
$G$ = $\langle g ^{p} \rangle$, $g$ as generator and $p$ is the
order of the group $G$ Common input: ($p,\,g$) Output: an element $k
\in G$ shared between Alice and Bob
\begin{enumerate}
  \item Alice chooses random number $a$ $\in U \otimes \textbf{1}$ and $p$, and
  send $g^a$ to Bob
  \item Bob: Choose random number $b$ between $1$ and $p$, and send
  $g^b$ to Alice
  \item Alice: compute $(g^b)^a$
  \item Bob: compute $(g^a)^b$
  \item By commutativity, Alice's $k_a$ = $g^{ba}$ = $g^{ab}$ =
  $k_b$. Notice that an adversary Eve intercepts {$g$, $g^a$, $g^b$}
  public information and cannot break the scheme with
  non-negligible probability. However this scheme is vulnerable to
  man-in-the-middle attack.
\end{enumerate}
\end{fmpage}
\end{center}


\subsection{Post-Quantum Public Key Exchange} \label{sect:QKE}

Public key cryptosystems and related protocols have been constructed
on the Turing machine model. The underlying theories are based on
\emph{Church-Turing's thesis}, which asserts that any reasonable
computation can be efficiently simulated on a probabilistic Turing
machine. New model of computing, quantum computation, has been
investigated since $1980$. Two most successful results are Shor's
probabilistic polynomial time algorithms for integer factorization
and discrete logarithm in the quantum Turing machine (QTM)
model\cite{Shor97} and Grover's unstructured search method
in$\sqrt{N}$ \cite{Grover96}. Although Shor's result demonstrates
the power of QTMs, Bennett, Bernstein, Brassard, and
Vazirani\cite{BBBV97} show that relative to an oracle chosen
uniformly at random, with probability $1$, class $NP$ cannot be
solved on a $QTM$ in time $O(2^{n/2})$. Many researchers consider
that it is hard to find a probabilistic polynomial time algorithm to
solve an $NP$-complete problem even in the QTM model.

Since Shor's result and Grover's search algorithm reduced many
practical public-key cryptosystems (RSA, multiplicative
group/elliptic curve versions of Diffie-Hellman and ElGamal schemes)
to insecure status, we need a quantum public-key cryptosystem
(QPKC). Many public key schemes such as BB84 and B92 were studied.
In 2000, Okamoto, \emph{et al} \cite{OTU00} proposed a theoretical
paradigm of QPKC that consist of quantum public-key encryption
(QPKE) and quantum digital signature (QDS). In our studies of
quantum channel and satellite communication, we realize an extension
of QPKC model and construct two practical schemes that achieve key
agreement. We discuss the possible attack and countermeasure of our
schemes.

If Eve has a quantum computer, she can easily break the logarithm
and get $a$ and $b$, then the secret key $((g^b\,\, \mbox{mod}\,\,
p)^a$ mod $p$).

The protocol of the Post-quantum Diffie-Hellman Key Exchange is
described below:

\begin{center}
\begin{fmpage}{15cm}

\textbf{Quantum Public Key Exchange Scheme}

\begin{enumerate}
  \item Alice and Bob use a quantum clock to synchronize their clocks.
  \item When Alice sends the message to Bob, she publicly announces to Bob
that they will start to count the bits at time $t$. (Due to the
different distance, Bob knows when he will start to count the bits
at time $t_1$). The key is $g$. They also agree on a prime number
$p$. $g$ and $p$ are public.
  \item  Alice teleports a quantum particle state to Bob and informs Bob that
she flips the $n^{th}$ bit of $g$. The position of the bit flipped
depends on the quantum state teleported by Alice to Bob. So both
Alice and Bob have the new key called $g_1$.
  \item Alice and Bob choose their secret keys $a$, and $b$, respectively. Alice
sends Bob ($(g_{1})^{a}$ (mod $p$)), and Bob sends Alice
($(g_{1})^{b}$ (mod $p$)). Both Alice and Bob have arrived at the
same value (($(g_{1})^b$ mod $p$) $^a$ mod $p$) or (($(g_{1})^a$ mod
$p$) $^b$ mod $p$).
  \item The key vanishes after it is used on Alice and Bob's site.
\end{enumerate}
\end{fmpage}
\end{center}

Only $p$ is public, Eve could intercept $(g_{1})^{a}$ (mod $p$) and
$(g_{1})^{b}$ (mod $p$). All $a$, $b$ and $g_1$ are secret.  Eve
could not figure out the key even she has a quantum computer or this
would make it too hard for her to compute the secret key. See Figure
2.

   \begin{figure}
   \begin{center}
   \begin{tabular}{c}
   \includegraphics [height=10cm, width=15cm] {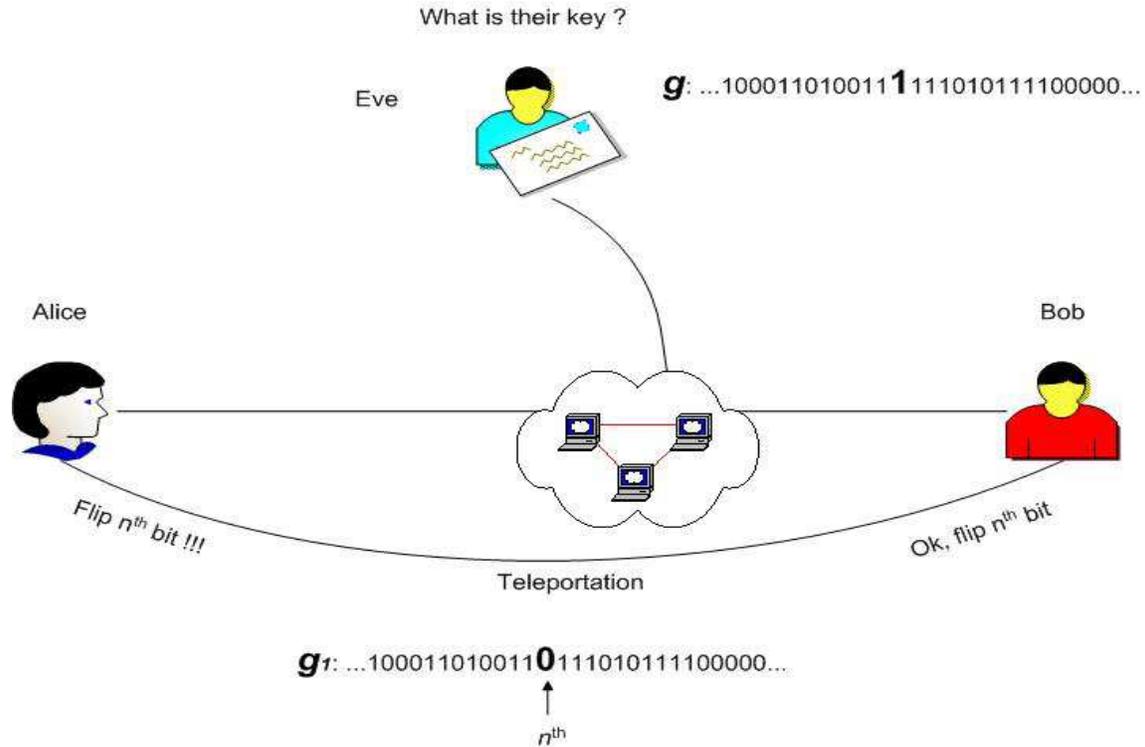}
   \end{tabular}
   \end{center}
   \caption[example]
   { \label{fig:example}
Post-quantum Diffie-Hellman Key Exchange.}
   \end{figure}

\subsection{Post-quantum Private Key Exchange Protocol}
The Private Key Encryption uses the same key to encrypt and decrypt
the message. Only Alice and Bob know the key. How do Alice and Bob
make the agreement on the key? They must trust the security of some
means of communications. Further, how do Alice and Bob secure the
key on their site? The key may be stolen.

The protocol of the post-quantum private key exchange is described
below:
\begin{center}
\begin{fmpage}{15cm}
\begin{enumerate}
  \item First, Alice and Bob use a quantum clock to synchronize their clocks.
  \item When Alice sends the message to Bob, she teleports a quantum
particle state to Bob. Both of them understand they will start to
count the bits at time $t$, (due to the different distance, Bob
knows when he will start to count the bits at time $t_1$).
  \item The key vanishes after it is used by Alice and Bob.
\end{enumerate}
\end{fmpage}
\end{center}

Eve could not get this entangled information, so she does not know
when Alice and Bob start to count the bits. Even Eve is at the same
site of Alice or Bob, she could not get the key since it disappears
after it is used. The keys used in encoding and decoding are used
once and are never stored.

\section{Quantum Random Walk Protocol}
In this section, we look at the quantum key distribution problem
under a slightly different consideration. We assume both Alice and
Bob have a simple quantum device, whereas Eve has a quantum
computer. Since the seminar work of BB84 and B92, quantum key
distribution (QKD) receives widespread attention because its
security is guaranteed by the law of physics and is different from
the classical counterparts\cite{Mao04}. Our scheme based on the
experimental realization\cite{Gisin02} and security
proof\cite{Mayers01,LoChau99,SP00} extends the KKKP
scheme\cite{Kye05} in two ways.

The procedure for the proposed quantum protocol is as follows:
\begin{center}
\begin{fmpage}{15cm}
\textbf{Quantum Walk Agreement Protocol}

\begin{enumerate}
  \item Alice and Bob perform a random walk on the random bits. In
  order to get an agreement, they both must use the same operator.
  \item Alice and Bob teleport or synchronize with a quantum clock to exchange the ``operator"
  \item Once they are in synchronization with the same operator, they apply the
  ``operator" on random bits stream, i.e. tree-walk the graph.
  \item Alice and Bob yield to the same key, i.e. the path of the operator-oriented walk on the graph.
\end{enumerate}
\end{fmpage}
\end{center}

Security of our scheme which minimizes the common problem of high
transmission rate of errors and defeats man-in-the-middle attack is
cleverly directed by quantum walk on one $q-bit$. Once quantum walk
determines the $q-bit$, Alice and Bob can use the agreed operator to
perform classical tree-walking on the random bits stream and
determine the key efficiently. Our scheme can be applied to any
quantum device that satisfies the above requirement.

In a similar vein, we formulate a quantum walk on a graph. Consider
a spin$-1/2$ particle that shifts to left or right depending on its
$spin$ state. Let a set of orthonormal basis states correspond to
vertices of the graph. If a particle is in the state $\ket{g}$, that
corresponds to a vertice $g$. (Another name for this technique is
commonly used by computational group theorists to carry argument
through for {\em Cayley-graphs} of Abelian groups and infinite
groups.)

We will look at the possible attacks from Eve's perspective. Eve
with a quantum computer can intercept all messages and perform a
quantum walk search\cite{AAKV01,Hillery03}. Our procedure modifies
the discrete quantum random walk result\cite{Kempe03} with a
different quantum device. Our quantum walk $U$ searches the graph
$G$ as follows:

\begin{center}
\begin{fmpage}{15cm}
\textbf{Quantum Walk Agreement Search Algorithm}
\begin{enumerate}
\item Initialize the quantum system in the uniform superposition $|\Phi_0\ra$.
\item Do $T$ times:
 Apply the marked walk $U'$.
\item Measure the position register.
\item Check if the measured vertex 
is the marked item.
\end{enumerate}
\end{fmpage}
\end{center}

The physical attack is that Eve can place a beam splitter attack
between the quantum channel and amplify the error rate.  Another
attack is by eavesdropping with phase shifters. Once Eve has an
estimate on the state pulse of quantum state, she can perform
probabilistic search of the key space $N$, i.e. the random bits
stream, on a $\sqrt{N}\times\sqrt{N}$ grid in time $O(\sqrt N \log
N)$ (See \ref{sect:qrw} for definitions).

Since the quantum walk search is restricted by initial condition and
localization of the quantum walk search, Eve is not guaranteed to
find the key in timely fashion for practical purpose.

\section{Conclusion}
We have shown that our schemes are secure against weak impersonation
attack, and quantum eavesdropping attacks. For future research on
quantum key agreement protocol, we like to consider the potential
weakness of random source generation on the satellite and carry out
experiment on the Elliptic pseudo random generation functions. One
open question is whether it is possible to extend our schemes with
the additional capability of entity authentication and signature?
For example, currently we are looking at the challenge of designing
quantum cryptographic voting protocols\cite{OST04}.

Another line of research undertaken by us investigates whether
quantum computer based on topological quantum
computation\cite{Kauffman05} with Anyons\cite{Collins06} and quantum
knots is easier to build and perform faster. Current schemes of
designing quantum computers use techniques to control interference
of quantum system with the ambient environment and lower the error
rates. As an alternative approach to the open problems of quantum
circuit complexity \cite{Yao93}, what can we say about braiding
operator\cite{Kauffman04} as the universal quantum gates?

On the quantum search problems, we are looking to extend the quantum
random walk techniques to arbitrary graphs, i.e. independent of
initial condition and localization problems and provide a better
bound on time and space.

\section{Acknowledgements}
It is a pleasure to thank Professor Mark Hillery for a discussion on
the quantum random walk subject.\\

\appendix    
\section{MISCELLANEOUS fundamentals} \label{sect:misc}

\subsection{Discrete Logarithm Problem} \label{sect:dlp}
We let $G$ = $\langle$ $g ^{n}$ $\rangle$ be a cyclic group
generated by $g$. By repeated squaring method, it is easy to compute
$g^{n}$ in $O$(log ${n}$) steps. Finding $n$ from $g$ and $g^n$ is a
hard problem with exponential complexity. The degree of
computational complexity depends on the representation of the group.
More generally in group-theoretic setting, given an isomorphism of
two finite group $G_1$, and $Z_k$ for $k \in N$, finding the image
of an element $g_n$ under the isomorphism map is equivalent to
solving the discrete log problem. A large variety of groups are
studied for use in the discrete logarithm problem.
\begin{enumerate}
  \item Subgroups of $Z_P$ for some prime $p$.
  \item Subgroups of $F_{p^n}$ for prime $p=2$
  \item Cyclic subgroups of the group of an elliptic curve
  $E_{a,b}(F_p)$ over the finite field $F_p$ with
  \begin{equation}\label{EC}
    Y^2 = X^3 + ax + b, \,\,\,\,\, a,b \in F_p
  \end{equation}
  \item The natural generalizations of the group of an elliptic
  curve to the Jacobian of a hyperelliptic curve
  \item Ideal class group of an algebraic number field
\end{enumerate}

A rigorous and formal security analysis with syntactical and
semantical consideration is in here\cite{Mao04}.

\subsection{Random Resource}

Most computer programming languages could generate random numbers.
In \emph{Lisp} the expression $\emph{(random 100)}$ produces an
integer in the range between $0$ and $99$, with each of the $100$
possible values having equal probability. But these are
pseudo-random numbers: They ``look" random, but under the surface
there is nothing unpredictable about them\cite{Hay01}.

The only source of true randomness in a sequence of pseudo-random
numbers is a ``seed" value that gets the series started. If you
supply identical seeds, you get identical sequences; different seeds
produce different numbers. The crucial role of the seed was made
clear in the 1980s by Blum. He pointed out that a pseudo-random
generator does not actually generate any randomness; it stretches or
dilutes whatever randomness is in the seed, spreading it out over a
longer series of numbers like a drop of pigment mixed into a gallon
of paint.

For most purposes, pseudo-random numbers serve perfectly well often
better than true random numbers. Almost all \emph{Monte Carlo} work
is based on them. Nevertheless, true randomness is still in demand,
if only to supply seeds for \emph{pseudo-random} generators. Finding
events that are totally patternless turns out to be quite difficult.

A obvious scheme for digitizing noise is to measure the signal at
certain instants and emit a $1$ if the voltage is positive or a $0$
if it is negative. Another popular source of randomness is the
radioactive decay of atomic nuclei, a quantum phenomenon that seems
to be near the ultimate in unpredictability.

Next we show an algorithm\cite{AG97} that achieves excellent uniform
distribution on seed generation.

\subsection{Random Number Generator and Elliptic-Zeta function} \label{sect:rngez}

Random number generator is an important mathematical tool. Van Dam
\cite{Dam05} shows that many known hard computational problems can
be exploited and solved by quantum factoring method and quantum
search algorithm, e.g Gauss Sums over finite rings. We have not seen
work that reduces Elliptic-Zeta function to Gauss Sums estimation.
We will reproduce definitions and theorems from the Anshel and
Goldfeld paper\cite{AG97} and describe three candidates of one-way
functions $F_{Kronecker}, F_{Elliptic}$, and $F_{Artin}$.

\subsubsection{Pseudorandom Number Generator.}

We adopt the notion of a pseudorandom generator suggested and
developed by Blum and Micali and Yao. A pseudorandom number
generator is a deterministic polynomial time algorithm that expands
short seeds into longer bit sequences such that the output of the
ensemble is polynomial-time indistinguishable from a target
probability distribution. We shall present an algorithm for a
cryptographically secure pseudorandom number generator that is based
on the candidate one-way function for the class
$\mathcal{Z}_{Elliptic}$, and $\mathcal{Z}_{Artin}$. We shall call
this pseudorandom number generator PNG$_{Elliptic}$. It has the
property that it transforms a short seed into a long binary string
of zeros and 1s with the target probability (1/3, 2/3) (i.e., the
probability of zero appearing is 2/3 while the probability of a 1 is
1/3). The proofs of these assertions are based on Theorems below.

    \textbf{Definition.} Let $\mathcal{P}$ be a set of primes having a certain property.
We define the density of $\mathcal{P}$ to be \[ \lim_{x\rightarrow
\infty} \sum_{p\in\mathcal{P},p\leq x} 1/\sum_{p\leq x}1,\] provided
the limit exists. If the limit does not exist, then the density of
$\mathcal{P}$ is not defined.

    With this definition, we now propose the following theorems.

\,\,\,\, \textbf{THEOREM 1.} Let $a,b$  determine an elliptic curve
$E:y^2=x^3+ax+b$. Define $d$ to be the degree of the field obtained
by adjoining the roots of the cubic equation $x^3+ax+b=0$ to
$\mathbb{Q}$. If $d$ =1,2, then $c_E(p)$ will be even for all except
finite many rational primes $p$. If $d$= 3, then the density of
primes for which $c_E(p)$ is even is 1/3 while if $d$=6, the density
is 2/3.

\,\,\,\, \textbf{THEOREM 2.} (Chebotarev). Let $K$ be a finite
Galois extension of $\mathbb{Q}$ with Galois group
$G=$Gal$(K/\mathbb{Q})$. For each subset $H\subset G$ stable under
conjugation (i.e., $\sigma H\sigma^{-1}=H, \forall\sigma\in G$), let
\[ \mathcal{P}_H=\{p\in\mathbb{Q},\,prime\,|\,Fr_p\in H\,
and\,\,p\,\,unramified\,\,in\,\,K\}.\] Then $\mathcal{P}_H$ has
density $|H|/|G|$, where $|H|,|G|$ denote the cardinalities of
$H,G$, respectively.

\,\,\,\, \textbf{THEOREM 3.} Let $E$ be an elliptic curve defined
over $\mathbb{Q}$. Let $K$ denote the field obtained by adjoining
the 2-torsion points of $E$ to $\mathbb{Q}$. Then there exists an
entire Artin L-function \[ L_K(s)=\sum_{n=1}^{\infty}b(n)\cdot
n^{-s} \in \mathcal{Z}_{\mbox{Artin}}\] of $K$ with the property
that \[ b(p)\equiv c_E(p)\,\,\,(\mbox{mod}\,2)\] for all except
finitely many rational primes $p$.

\subsubsection{Coin Flipping by Telephone.}

Alice and Bob want to simulate a random coin toss over a telephone.
The following algorithm provides a mechanism for accomplishing this
task. The algorithm assumes that $B\rightarrow \infty$ and $m=(\log
B)^k$ for some constant $k>2$.

\,\, $Step \,\,1.$ Alice chooses integers $a, b$ such that the roots
of the equation $x^3+ax+b=0$  generate a field of degree 6 over
$\mathbb{Q}$, and the discriminant $\triangle =4a^3+27b^2$ lies in
the interval $B\leq\triangle\leq2B$. Alice then computes the vector
$v$ of the first $m$ coefficients \[ v=\{a(1),a(2),...,a(m)\}\] of
the Zeta function associated to $E:y^2=x^3+ax+b$. Alice transmits
$v$ to Bob.

\,\, $Step \,\,2.$ Bob randomly chooses two prime numbers $p<p'$
with $p>m$.

\,\, $Step\,\, 3.$ Alice computes trial $(p,p^{'})$=(a(p) (mod 2),
$a(p')$ (mod 2)). If \[ \mbox{trial}(p,p') = (1,0),\] then the coin
toss is heads. If \[ \mbox{trial}(p,p') = (0,1),\] then the coin
toss is tails. If neither of these possibilities occur, go back to
Step 2.

\,\, $Step \,\,4.$ Bob can verify the correctness of the coin flip
when Alice announces the elliptic curve $E$. Otherwise it is not
feasible for him to compute trial $(p,p')$.

\,\,The probability of either of the events, trial $(p,p')$ =(1,0)
or (0,1), is 2/9, so they will occur with equal frequency.

\subsection{Quantum Teleportation}

Quantum teleportation (QT) \cite{Collins06,Kauffman04,Kauffman05} is
a particularly attractive paradigm. It involves the transfer of a
quantum state over an arbitrary spatial distance by exploiting the
prearranged entanglement (correlation) of ``carrier" quantum systems
in conjunction with the transmission of a minimal amount of
classical information. This concept was first discussed by Aharonov
and Albert\cite{Aharonov81} (AA) using the method of nonlocal
measurements.

Over a decade later, Bennett, Brassard, Crepeau, Jozsa, Peres, and
Wootters (BBCJPW)\cite{BBCJPW93} developed a detailed alternate
protocol for teleportation. It consists of three stages. First, an
Einstein-Podolsky-Rosen (EPR) \cite{EPR35} source of entangled
particles is prepared. Sender and receiver share each a particle
from a pair emitted by that source. Second, a Bell-operator
measurement is performed at the sender on his \emph{EPR}
\cite{EPR35} particle and the \emph{teleportation}-target particle,
whose quantum state is unknown. Third, the outcome of the Bell
measurement is transmitted to the receiver via a classical channel.
This is followed by an appropriate unitary operation on the
receiver's \emph{EPR} particle. To justify the name
``\emph{teleportation}"\cite{BBCJPW93}, notice that the unknown
state of the transfer-target particle is destroyed at the sender
site and instantaneously appears at the receiver site. Actually, the
state of the \emph{EPR} particle at the receiver site becomes its
exact replica. The teleported state is never located between the two
sites during the transfer.

The first laboratory implementation of QT was carried out in $1997$
at the University of Innsbruck by a team led by Anton
Zeilinger\cite{Bouwmeester97}. It involved the successful transfer
of a polarization state from one photon to another.

\subsection{Quantum Clock Synchronization} Clock
synchronization\cite{Simons90,LAK99} is an important problem with
many practical and scientific applications. Alice and Bob, both have
good local clocks that are stable and accurate, and wish to
synchronize these clocks in their common rest frame. The basic
problem is easily formulated: determine the time difference
\emph{$\Delta$} between two spatially separated clocks, using the
minimum communication resources\cite{Chuang00}. Generally, the
accuracy to which \emph{$\Delta$} can be determined is a function of
the clock frequency stability, and the uncertainty in the delivery
times for messages sent between the two clocks. Given the stability
of present clocks, and assuming realistic bounded uncertainties in
the delivery times, protocols have been developed which presently
allow determination of \emph{$\Delta$} to accuracies better than 100
$ns$ (even for clock separations greater than $8000$ $km$); it is
also predicted that accuracies of 100 $ps$ should be achievable in
the near future.

A quantum bit (\emph{q-bit}) behaves naturally much like a small
clock. For example, a nuclear spin in a magnetic field processes at
a frequency given by its gyromagnetic ratio times the magnetic field
strength. And an optical \emph{q-bit}, represented by the presence
or absence of a single photon in a given mode, oscillates at the
frequency of the electromagnetic carrier. The relative phase between
the $|0\rangle$ and $|1\rangle$ states of a \emph{q-bit} thus keeps
time, much like a clock, and ticks away during transit. Unlike a
classical clock, however, this phase information is lost after
measurement, since projection causes the \emph{q-bit} to collapse
onto either $|0\rangle$ or $|1\rangle$, so repeated measurements and
many \emph{q-bits} are necessary to determine \emph{$\Delta$}. On
the other hand, with present technology it is practical to
communicate \emph{q-bits} over long distances through
fibers\cite{Hughes95,MZG96}, and even in free space\cite{Buttler98}.

Let $t^{a}$ and $t^{b}$ be the local times on Alice and Bob's
respective clocks. We assume that their clocks operate at exactly
the same frequency and are perfectly stable. The goal is to
determine the difference \emph{$\Delta$}= $t^{b}$ - $t^{a}$, which
is initially unknown to either of them. Quantum
synchronization\cite{Burgh98,Valencia04,Janzing03} comes in many
schemas. Chuang\cite{Chuang00} accomplished this goal by using the
Ticking qubit handshake (\emph{TQH}) protocol. He also established
an upper bound on the number of \emph{q-bits} which must be
transmitted in order to determine \emph{$\Delta$} to a given
accuracy.  Chuang found that only $\emph{O(n) q-bits}$ are needed to
obtain $n$ bits of \emph{$\Delta$}, if we have the freedom of
sending \emph{q-bits} which tick at different frequencies.

\subsection{Quantum Random Walk} \label{sect:qrw}

We provide the standard notation\cite{CLR90} and model on tree and
graph for our discussion. Then we also briefly describe the
connection between the coined quantum random walk and the graph
representation of quantum state. We cite the main result of quantum
random walk theorem\cite{Kempe03} used in our arguments.

Let us look at an example of tree $T(V,E)$ that consists of vertices
and edges. Consider a $3-bits$ binary tree \emph{T}. $T$ with depth
of $3$, has ${2^{3} = 8}$ (vertices) binary numbers represented at
its leaves. The topmost level of \emph{T} is denoted `\emph{root}'
and the bottom level of \emph{T} is denoted \emph{leaf}. The
prefixes associated with subtrees are denoted in italics. In this
example, we consider three leaves the '\textit{001}','\textit{011}',
and '\textit{110}' vertice. The tree-walking algorithm, a recursive
depth first algorithm, here first singulates the '\emph{001}' leave.
It does this by following the path that connects two vertices and
the complexity is ${O(\log n)}$.

Formally, given an {\em undirected graph} $G=(V,E)$ that each vertex
$v$ stores a variable $a_v\in\{0, 1\}$, our goal is to find a vertex
$v$ for which $a_v=1$ (assuming such vertex exists). We will often
call such vertices marked and vertices for which $a_v=0$ unmarked.

In one step, an algorithm can examine the current vertex or move to
a neighboring vertex in the graph $G$. The goal is to find a marked
vertex in as few steps as possible.

A quantum algorithm is a sequence of unitary transformations on a
Hilbert space\cite{AAKV01}\,. $\H_i\otimes \H_V$. $\H_V$ is a
Hilbert space spanned by states $\ket{v}$ corresponding to vertices
of $G$. $\H_i$ represents the algorithm's internal state and can be
of arbitrary fixed dimension. A $t$-step quantum algorithm is a
sequence $U_1$, $U_2$, $\ldots$, $U_t$ where each $U_i$ is either a
{\em query} or a {\em local transformation}. A query $U_i$ consists
of two transformations ($U_i^0$, $U_i^1$). $ U_i^0  \otimes I$ is
applied to all $\H_i \otimes |v\ra $ for which $a_{v}=0$ and $ U_i^1
\otimes I$ is applied to all $\H_i \otimes |v\ra $ for which
$a_{v}=1$.

A local transformation can be defined in several ways. In this
paper, we require them to be $Z$-local. A transformation $U_i$ is
$Z$-local if, for any $v\in V$ and $\ket{\psi}\in \H_i$, the state
$U_i(\ket{\psi}\otimes \ket{v})$ is contained in the subspace
$\H_i\otimes \H_{\Gamma(v)}$ where $\H_{\Gamma(v)} \subset \H_V$ is
spanned by the state $\ket{v}$ and the states $\ket{v'}$ for all
$v'$ adjacent to $v$. Our results also apply if the local
transformations are $C$-local.

The algorithm starts in a fixed starting state $\ket{\psi_{start}}$
and applies $U_1$, $\ldots$, $U_t$. This results in a final state
$\ket{\psi_{final}}=U_t U_{t-1} \ldots U_1 \ket{\psi_{start}}$.
Then, we measure $\ket{\psi_{start}}$. The algorithm succeeds if
measuring the $\H_V$ part of the final state gives $\ket{g}$ such
that $a_{g}=1$.

\,\,\,\, \textbf{THEOREM\cite{Kempe03} 4.} The associated quantum
walk search algorithm takes $O(\sqrt{N \log N})$ steps  and the
probability to measure the marked state is $\Omega(1/\log N)$. This
yields a local search algorithm running in time $O(\sqrt{N}\log N)$.


}






\end{document}